# Software Cognitive Complexity Measure Based on Scope of Variables

Kwangmyong Rim and Yonghua Choe
Faculty of Mathematics, Kim Il Sung University, D.P.R.K
mathchoeyh@yahoo.com

**Abstract**
In this paper, we define a Mathematical model of program structure. Mathematical model of program structure defined here provides unified mathematical treatment of program structure, which reveals that a program is a large and finite set of embedded binary relations between current statement and previous ones. Then, a program is considered as a composed listing and a logical combination of multiple statements according to the certain composing rules. We also define the Scope Information Complexity Number (SICN) and present the cognitive complexity based on functional decomposition of software, including theoretical validation through nine Weyuker's properties.

## 1 Introduction

Cognitive complexity measures attempt to quantify the effort or degree of difficulty in comprehending the software based on cognitive informatics foundation that "cognitive complexity of software is dependent on three fundamental factors: inputs, outputs, and internal processing".

In 2003, Cognitive Function Size (CFS) [2] was suggested and satisfied 8 properties of Weyuker, then, many approaches [3-9] have been modified from "Cognitive Function Size" (CFS) to fully consider complexity factors. In 2009, [1] selected the complexity factors more analogue in human understanding and suggested new complexity metric.

As cognitive complexity measures attempt to consider all factors affecting the effort in comprehending the software, e.g. loops and branches, data objects such as inputs, outputs, and variables, evaluating complexity from many factors can be troublesome if the factors are not carefully thought and organized. In [7-9], that process of calculation ignores relationships among the factors and has little relevance to human cognitive process when comprehending the program code.

Mathematical model of Program [10] is a model for showing executive step of Program. Information complexity number of variable depends on the value change of variable and cognition of its value change depends on the scope of variable. They defined information complexity number of variable and information of Program based on assumption that variables and operators contained information of [2]. But cognition of Program is cognition of function, class, module and file in cognition and understanding of Program, they implement with source code and final it is cognition of source code.

In this paper, we define scope information complexity of variables to understand the meaning of each variable according to the scope, suggest a mathematical model of Program for program complexity and a cognitive complexity metric based on relation between scopes.

## 1 Scope Information Complexity Number of Variables

In [1], information complexity number of variables(ICN) is followed by:
> *At the beginning of the program, the Informatics Complexity Number (ICN) of every variable* is *zero. When a variable* is *assigned the value in the program, its ICN increases by I, and if that assignment statement contains operators, ICN of the variable that* is *assigned the value also further increases by the number of operators in that statement.*

*For variable 'V' , L is a program,* $ICN_{max}(V, L)$ *is the highest ICN of V 's occurrences in L.I(L) is defined as the sum of* $ICN_{max}(V, L)$ *of every variable V exists in L.*

**Example 1.**
```
public static void main(string[] args)
{ int UserInput;
   int square;
      UserInput=Text1.getInt();
      Square=UserInput*UserInput;
      System.out.print(square);
}
```

In Example 1,
ICN(UserInput)=1
ICN(square)=2
I(L)=1+2=3

Cognition of variables based on the scope of variables. When a variable is used as private variable and public variable, the variable is used as different function in different action scope, therefore its meaning of information is not same.

**Example 2.**
```
int amount=123;//public variable
amount=amount*2;
void main()
{
    int amount=456;//local variable
    amount=amount+1;
    cout<<::amount;// public variable
```

```
        {
                int amount=789;//other local variable
                amount=amount--;
                cout<<::amount;// public variable that is out of method main
                                //variable is not 456
                cout<<amount;//amount is local variable 789
        }
}
```
  In Eg 2, amount variable is public variable as well as local variable. According to variable's scope, its meaning is different and it must be comprehended each case. In order to measure more cognitive and comprehensive complexity with the scope of variables and decomposition methodology of BCS unit of software, we define the concept of scope of variables and scope information of program followed by:

**[Definition 1]** *Scope Information Cognitive Number of Variables ( SICN )*
1) SICN of single variable
   SICN of variable of program equals its ICN in the same variable scope.
   If variable is not only public variable but private variable, we recognize different variables and calculate their SICN in each scope.
2) ) SICN of structure variable

SICN of structure variable is a sum of SICNs of member variables.

**[Definition 2]** *L is program or part of program.*
 *For variable 'V' appearing in L, $SICN_{max}(V,L)$ is the highest SICN of V's occurrences in L.*
*For variable 'V' appearing in L, $SICN_{min}(V,L)$ is the lowest SICN of V's occurrences in L.*

**[Definition 3]** Scope Information contained in L (SI(L))
*Scope Information contained in L (SI(L)) is defined as the sum of $[SICN_{max}(V,L) - SICN_{min}(V,L)]$ of every variable V exists in L.*

$$SI(L) = \sum_{V \in L}(SICN_{max}(V,L) - SICN_{min}(V,L))$$

*Scope Information* is defined value change number of variable in some scope to express cognition of variable. Cognition Information of program '*L*' is value change number of variable in *L*.

**Example 3**
```
#include<ostream.h>
void main()
{    int key[]={20,10,50,40,60,70,30,45,67,15};
     int n=10,s=0;
     int left=1,right=n,m=n,buf;
```

```
    for(int i=0;i<n;i++)
    s=s+key[i];
    while(left<right)
    { int i=n;
         while (i>0)
         {    if(key[i-1]>key[i])
              {   buf=key[i-1];
                  key[i-1]=key[i];
                  key[i]=buf;
              }                                              L1
              m=i;
              i=i-1;
         }
left=m+1;
for(int s=left;s<=right;s++)
{
    if(key[s-1]>key[s])
    {   i=i+key[s];
        buf=key[s-1];
        key[s-1]=key[s];
        key[s]=buf;
        s=s-1;                                               L2
    }
    m=s;
}
    right=m-1;
}
for(int i=0;i<n;i++)
    cout<<key[i]<<"\t";
cout<<::s;
}
```

In Example 3, this shows comparison of $ICN$ and $SICN$.

$ICN_{\max}(s, L1) = 3$

$SICN_{\max}(s, L1) = 3$

$ICN_{\max}(s, L2) = 8$

$SICN_{\max}(s, L2) = 5$

$W_{BCS}(L1) = 4, W_{BCS}(L2) = 4$

• Complexity by $ICN$
  $W_{BCS}(L1) * I(L1) + W_{BCS}(L2) * I(L2) = 44$

• Complexity by $SICN$
  $W_{BCS}(L1) * SI(L1) + W_{BCS}(L2) * SI(L2) = 32$

## 3 Mathematical model of Program for complexity

Mathematical model of Program[10] is a model for showing executive step of Program.

**[Definition 4]** $BCS_p$ is a set of Basic control structures in program following as;

$$BCS_p = \{@, \langle, \frown, \bot, R^i, R^+, R^*, \xrightarrow{f}, \circlearrowright \}$$

**Table 1    Basic control structures**

| No | BCS | Sign | Implementation |
|---|---|---|---|
| 1 | Simple | $@$ | assignment |
| 2 | Goto | $\frown$ | go to |
| 3 | branch | $\langle$ | if |
|   |        | $\bot$ | select case |
| 4 | Iterative | $R^i$ | while |
|   |           | $R^+$ | do-while |
|   |           | $R^*$ | for |
| 5 | Function call | $\xrightarrow{f}$ | |
| 6 | Recursive | $\circlearrowright$ | |

**[Definition 5]** $R_{BCS}$ is a set of relation of BCSs in program following as;

$$R_{BCS} = \{\rightarrow, \supset, \leftrightarrow\} \quad r_1 : \rightarrow \text{ :Sequence}, \quad r_2 : \supset \text{ :Include}, \quad r_3 : \leftrightarrow \text{ :Concurrency}$$

**Table 2    Control structure relations and their models**

| Relation | Sign | Model | Example |
|---|---|---|---|
| Seq | $\rightarrow$ | 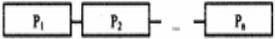 | $P_1 \rightarrow P_2 \rightarrow \cdots \rightarrow P_n$ |
| Inc | $\supset$ | 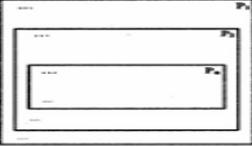 | $P_1 \supset P_2 \supset \cdots \supset P_n$ |
| Con | $\leftrightarrow$ | 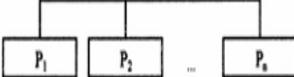 | $P_1 \leftrightarrow P_2 \leftrightarrow \cdots \leftrightarrow P_n$ |

**Example 4.**

void main( )

{

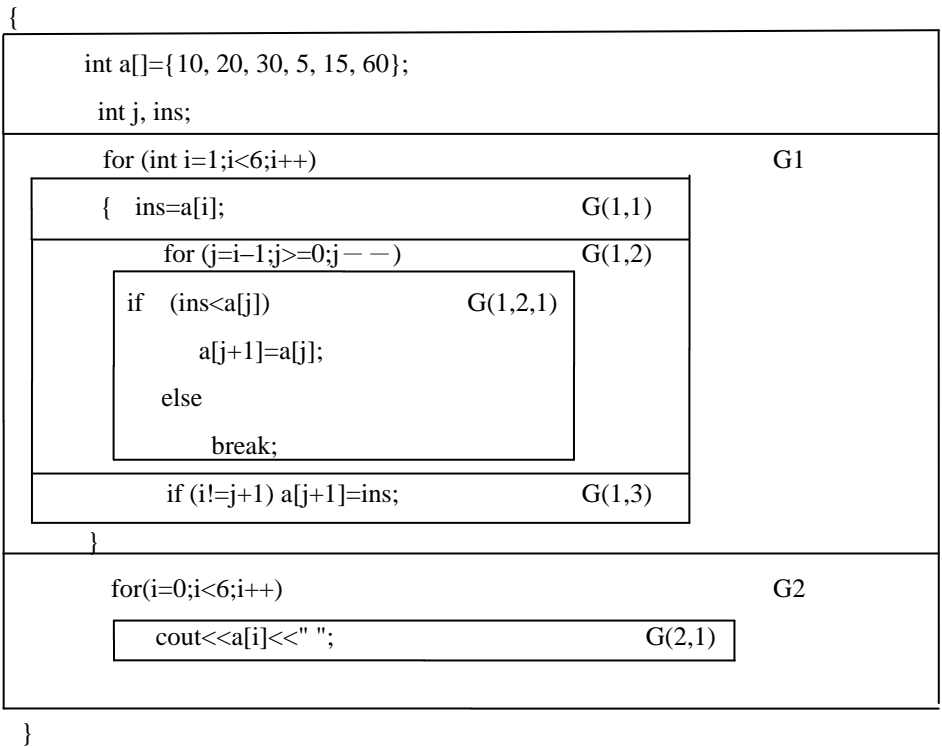

}

G1 → G2,
G1 ⊃ G(1,1), G1 ⊃ G(1,2), G1 ⊃ G(1,3)
G(1,1) → G(1,2), G(1,2) → G(1,3)
G(1,2) ⊃ G(1,2,1)
G(2) ⊃ G(2,1)

    Program is a set of finite lists of statements and each statement can be showed by basic control structures. Program is a set of statements which carry out function by variance of variables, so we have to consider function and scope of variables, and then we can suggest a complexity metric which can measure cognitive complexity of program by considering scope and variance of variables. The scope of variable is determined by declaration of variable.

    Program is a set of declaration of variables and statements, statements is combined by some relations. Structure of program is showed by variables, basic control structures and relations of control structures. Scope of variables is important to understand the function and consider variance of variables.

    And combination of statements can be represented by BCS and relation of control structures. Whole program can be represented by scope of variables and BCSs.

Now we can consider cognitive complexity of program in a viewpoint of structure and model a program for quantitative calculation of complexity.

**[Definition 6]** A program $\beta$ is a 4-tuple, i.e.:

$$\beta = (V, BCS_p, R_{BCS}, T)$$

V is a set of declaration of variables.
$BCS_P$ is a set of basic control structures(BCS).
$R_{BCS}$ is a set of relations of control structures.
T is a set of elements made by following rules.

t∈T:

① t→φ

② t→v(∈V): declaration section of variables

③ t→b(∈ $BCS_P$): basic control structures

④ t→v∘t: concatenation with declaration of variables

⑤ t→brt, (b,t∈ $BCS_P$, r∈ $R_{BCS}$, brt:=b∪t) (r is relation between b and t)

⑥ t→b∘t, t∈V, b∈ $BCS_P$

if $C_i \in V$ then $r_{ij}$ is concatenation '∘' else $r_{ij} \in R_{BCS}$, $j = i+1$
n is number of BCSs and declaration of variables of T
We call t(∈T) control structure.

**[Theorem]** Any Control structure t(∈ T) is represented finite lists of BCSs and declaration of variable by relations of Control structure and its representation is unique. In other words: $\forall t \in T$,

$$t = \underset{i=1}{\overset{n-1}{R}} C_i r_{ij} C_j = C_1 r_{12} C_2 r_{23} C_3 \cdots r_{n-1,n} C_n \quad (1)$$

$C_i, C_j \in \{BCS_p, V\}$,

Proof.  $\forall t \in T$
  ① $t \in BCS_p \Rightarrow t = (C)$, ($C \in BCS_p$, n=1)
  ② $t \notin BCS_p \Rightarrow t \in T \setminus BCS_p$
     $\exists! a, b \in T, \exists! r \in R_{BCS}; t = arb$
   if $a \in BCS_p \wedge b \in BCS_p \Rightarrow T = C_1 r C_2$, $C_1 = a, C_2 = b$
$a \notin BCS_p \vee b \notin BCS_p \Rightarrow \exists! a', a1', b', b1' \in S, \exists! r' \in R_{BCS}$
$(a = a' r' b') \vee (b = a1' r' b1')$
If $C_i$ is in V then t is t∘ $C_i$.

And then we continue in same ways, we will get following results.

$$t = \underset{i=1}{\overset{n-1}{R}}(C_i r_{ij} C_j) = C_1 r_{12} C_2 r_{23} C_3 \cdots r_{n-1,n} C_n$$

$$C_i, C_j \in \{B\ C\ S_p, V\}$$

If $C_i \in V$ then $r_{ij}$ is concatenation '∘' else $r_{ij} \in R_{BCS}$, $j = i+1$

(End)

Control structure is a embedded relational set of BCSs and declaration of variables. The ERM model provides a unified mathematical treatment of programs, which reveals that a program is a finite and nonempty set of embedded binary relations between a current statement and all previous ones that formed the semantic context. Program is a set of finite lists of statements and each statement can be showed by basic control structures.

**[Corollary]** Program $\beta$ is a Embedded Relational Model by finite combination of BCSs and declaration of variables.

**Example 5.**
```
struct node
{   int i;
int j;
int value;
node*next;
}
const int M=5, N=5;//5*5 matrix
void Insert(node**h, int i, int j, int x)
{if(*h==0)
{*h=new node;
(*h)‐>i =i, (*h)‐>j=j, (*h)‐>value=x;
(*h)‐>next =0;
return;
}
node*p=0, *q=*h;
while(q && ((q‐>i ‐ 1)*M+q‐>j<(i−1)*M+j))
{ p=q;
q=q‐>next ;
}
node**temp=(p==0)?h:&p‐>next ;
*temp=new node;
(*temp)‐>i=i;
(*temp)‐>j=j;
(*temp)‐>value =x;
(*temp)‐>next =q;
```

```
}
_______________________________________________
void Delete(node*h, int i, int j)
{ if(h==0) return;
node*p=0, *q=h;
while(q && ((q - >i - 1)*M+q–>j!=(i - 1)*M+j))
{ p=q;
q=q - >next ;
}
if(!q) return;
p - >next =q - >next ;
delete q;
}
_______________________________________________
```

Program is a ERM of BCSs, complexity of program depends on the complexity of combination of BCSs. Therefore, we can suggest a complexity metric of program by combination of BCSs and scope of variables.

## 4 Cognitive Information Complexity of Software : ESCIM

### 4.1 Decomposition of Software into BCS Hierarchical Structure

In this section, we suggest extended Structural Cognitive Information Measure based on scope information complexity of variables and BCS unit decomposition of software. To apply granular computing strategies to cognitive complexity measurement, first we decompose software into a hierarchy of granules.

When we comprehend the software, a BCS can be seen as a comprehension unit of which we need to understand functionalities and inputs/outputs before understanding interaction between BCS units and the whole program. Therefore, in the context of cognitive complexity measurement, we view a granule as a basic control structure (BCS), which may contain nested inner BCS's and information content. The decomposition methodology of the program can be explained as followed:

1) At the top level of the hierarchy, the whole program is partitioned into granules of BCS' s in linear structure.

2) Each granule whose corresponding BCS contains nested BCS's inside, is further partitioned generating next level of hierarchy.

3) The partitioning stops when corresponding BCS to the granule is a linear BCS.

In brief, each level of the hierarchy consists of BCS's in linear structure, and because a BCS that contains no nested BCS's inside can be said to contain a single linear BCS, leaf nodes of the decomposed hierarchy are the linear BCS's. An example construction of the hierarchy from a program from [1] can be demonstrated as in Fig 1.

**Example 6.**
```
       public static void main(string[] args)
       {
         int []numbers;                          G1
         int numcount;
         int num;
         numbers=new int[100];
         numcount=0;
         Text1.putln("Enter 10 integers");
         while (true)                            G2
         {   Text1.putln();                      G(2,1)
             num=text1.getlnInt();
                   if (n<=0)                     G(2,2)
                       break;
             numbers(numcount)=num;              G(2,3)
             numcount++;
         }
         ………………
       }
```
Demonstration of BCS hierarchical structure construction of Eg 4 is followed by:

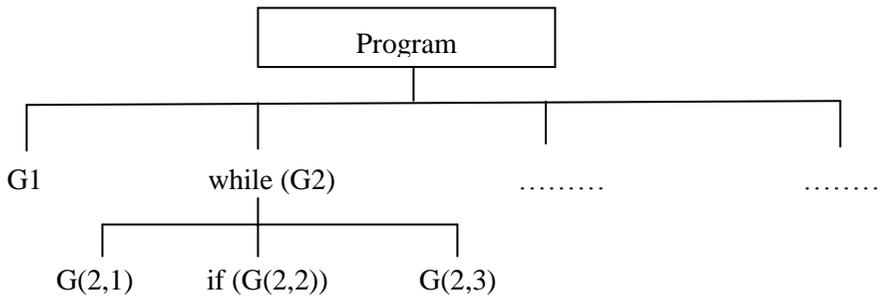

### 4.2 The Extended Structural Cognitive Information Measure of Software (ESCIM)

**[Definition 7]** ESCIM is defined as the total sum of the products of corresponding cognitive weights and information contained in leaf node granule (I(L)). Since software may consist of q linear blocks composed in individual BCS 's, and each block may consist of 'm' layers of nesting BCS's, and each layer with 'n' linear BCS 's, then

$$SI(t_i) = \sum_{V \in t_i}(SICN_{\max}(V,t_i) - SICN_{\min}(V,t_i))$$

$$C(t_i) = \prod_{k=1}^{m_i} W_c(j,k) * [\sum_{i=1}^{n_i} SI(j,k,i) * W_c(j,k,i)]$$

$$ESCIM(S) = \sum_{i=1}^{q} C(t_i)$$

where weights $W_c(j,k)$ of BCS's are cognitive weights of BCS's presented in [2], and $SI(j,k,i)$ are information contained in a leaf BCS granule as defined in Definition 3.

From the Definition, we can say that ESCIM evaluates the complexity by taking into account the dependencies of variables and their position in the BCS's structure as suggested by Fig 1.. Number of inputs/outputs can now be disregarded as 1I0s variables have already been included as the information contained in the program.

### 4.3 The Unit of ESCIM

In ESCIM, the simplest software component with only one variable assignment, no operators, and a linear sequential BCS structure, is defined as the Extended Structural Cognitive Information unit (ESCIU), computing ESCIM can be formulated as:

ESCIM= 1 * 1 = 1 [ESCIU]

The value in SSCU of a software system indicates its cognitive complexity relative to that of the defined simplest software component,

$$\text{ESCIU} = \frac{\text{cognitive complexity of the system}}{\text{cognitive complexity of the defined simplest software component}}$$

## 5 Validation Through Weyuker Properties

The proposed ESCIM can be proved to satisfy all nine Weyuker's properties, which are often used to evaluate and compare complexity measures as shown in Table 2.

Comparison of Conformance of Complexity Measures to Weyuker's Properties

|   | LOC | McCabe's Cyclomatic | Halstead's Effort | Dataflow Complex | CFS | MCCM | CPCM | SCIM | ESICM |
|---|---|---|---|---|---|---|---|---|---|
| 1 | / | / | / | / | / | / | / | / | / |
| 2 | / |   | / | × | / | / | / | / | / |
| 3 | / | / | / | / | / | / | / | / | / |
| 4 | / | / | / | / | / | / | / | / | / |
| 5 | / | / | × | × | / | / | / | / | / |
| 6 | × | × | / | / | × | × | × | / | / |
| 7 | × | × | × | / | / | × | × | / | / |
| 8 | / | / | / | / | / | / | / | / | / |
| 9 | × | × | / | / | / | / | / | / | / |

Let P and Q be program body.

**Property 1.** (∃ P)(∃ Q)(|P|≠|Q|)

This property states that the measures should not rank all the programs as equally complex. Therefore, ESCIM obviously satisfies this property.

**Property 2.** (∀ P),|P|≥0

Let c be a nonnegative number, then there are only finitely many programs of complexity c. Since all programming languages can have only finite number of BCS's, variable assignments, and operators, it is assumed that some largest numbers can be used as an upper bound on the numbers of BCS's, variable assignments and operators. Therefore, for these numbers, there are finite many programs having that much number of BCS's, variable assignment, and operators. Consequently, for any given value of ESCIU, there exists finitely large number of programs, and ESCIM satisfies this property.

**Property 3.** (∃ P)(∃ Q)(|P|=|Q|)

There are distinct program P and Q such that !PI =IQI. ESCIM clearly satisfies this property as at least for any program containing operator '+', replacing '+' with '-' will result in a different program with the same ESCIM complexity.

**Property 4.** (∃ P)(∃ Q)(P=Q & |P|≠|Q|)

This property states that there exist two programs equivalent to each other (i.e. for all inputs given to the program, they halt on the same values of outputs.) with different complexity. Clearly, the program computing 1+2+…+n can be implemented with while loop, or simply sequence structure with formula n(n+1)/2. The values of ESCIM from these two implementations are different. Hence, ESCIM satisfies this property.

**Property 5.** (∃ P)(∃ Q)(|P|≤|P;Q| & |Q|≤|P;Q|)

ESCIM obviously satisfies the property because adding any program body whether to the end or before the beginning of a program body can only increase or hold the ESCIM complexity.

**Property 6a.** (∃ P)(∃ Q) (∃ R) (|P|=|Q| & |P;R|≠|Q;R|)

Given program P and Q with same value in ESCIU, and program R contains some variables that are assigned values in P but no variables that are assigned values in Q, IP;RI is clearly more than IQ;RI because SICNs of those variables in R of P;R are higher than those of the same variables in R of Q;R. Therefore, ESCIM satisfies this property.

**Property 6b.** (∃ P)(∃ Q) (∃ R) (|P|=|Q| & |R;P|≠|R;Q|)

In the same way as in property 6a, ESCIM satisfies this property. The satisfaction of property 6 indicates one strength of ESCIM over other cognitive complexity measures that when different programs with the same complexity value are extended with the same program part, other measures view the extended programs as having the same complexity no matter what. This is because they do not consider possible complexity transferred between BCS in linear structures, or view linear BCS's as completely separately comprehensible, while ESCIM estimates the complexity transferred between blocks of BCS by the cumulative variable complexity counting scheme and does not overlook interrelationships among granules.

The intent behind Weyuker's Properties is to check whether complexity value of a program is suitable with complexity values of its parts. However the Definitions leave some room for measures to slip through. For example, CICM happens to satisfy Property 6 because its weighing of information content is so random that there exist programs P, Q, R that IPI=IQI but IP;RI $\neq$ IQ;RI.

Even though sometimes, if R is completely independent of P and Q, IP;RI should be the same as IQ;RI. We can say that the measure that truly satisfies the intent of Weyuker's properties should be able to answer what would happen to IP;RI when P and R are in some condition to each other. For ESCIM, IP;RI equals to IPI+IRI when cognition of R in IP;RI is not affected by P, while IP;RI > IPI+IRI when P has some effects on R.

**Property 7.** There are some program bodies P and Q such that Q is formed by permuting the order of statements of P, and |P|≠|Q|.
ESCIM satisfies this property because the permutation of statements can result in different SICNs, hence making the ESCIM value different.

**Property 8.** If P is renaming of Q, then IPI = IQI
ESCIM clearly satisfies this property as it does not take into account the names.

**Property 9.** ($\exists$ P)($\exists$ Q)(|P|+|Q|≤|P; Q|)
ESCIM satisfies this property because if some variables assigned values in P occur in Q, the complexity of Q in P;Q will increase from Q alone because the SICNs of those variable will increase, hence making IP;QI higher than IPI + IQI.
CFS,SCIM and ESCIM can indicate the coding efficiency (E), which can be defined as:

$$E = \frac{ESCIM}{LOC}$$

The higher coding efficiency indicates the higher complexity information packed in the shorter program code, therefore the program is likely to contain more defects than the program with lower coding efficiency.

**References**

[1] B.Auprasert and Y.Limpiyakorn,"Structuring Cognitive Information for Software Complexity Measurement" , Accepted for CSIE 2009,LosAngeles,USA,April 2009

[2] Y.Wang and J.Shao, "Measurement of the Cognitive Functional Complexity of Software", Proc. of 2nd IEEE Int'l Conference on Cognitive Informatics, p.67, Aug. 2003.

[3] D.S.Kushwaha and A.K.Misra, "A modified cognitive information complexity measure of software", ACM SIGSOFT Software Engineering Notes v.31 n.1, Jan.

2006.

[4] S.Misra, "Modified Cognitive Complexity Measure", Computer and Information Sciences – ISCIS 2006, pp. 1050-1059, Springer Berlin / Heidelberg, October 2006.

[5] S.Misra, "Cognitive Program Complexity Measure", Proc of 6th IEEE Int'l Conf on Cognitive Informatics, p.120, 2007.

[6] Wang Y, "On the Cognitive Informatics Foundations of Software Engineering", Proc. of 3rd IEEE Int'l Conference on Cognitive Informatics, 2004.

[7] YingXiu Wang," Cognitive Complexity of Software and its Measurement"5th,IEEE Int. Conf. on Cognitive Informatics(Iccro6),2006,p 226-235

[8] S.Misra and A.K.Misra," Evaluation and comparison of cognitive complexity measure" ACM SIGSOFT Software Engineering Notes v.32 n.2,Mar.2007

[9] S.Misra and A.K.Misra,"Evaluating cognitive complexity measure with Weyuker properties", Proc of the 3rd ICCI,16-17 Aug.2004

[10] Wang Y, "A UNIFIED MATHEMATICAL MODEL OF PROGRAMS", IEEE CCECE/CCGEI, Ottawa,2381~2384, May 2006.